# Protocols and Open Issues in ATM Multipoint Communications [1]

Sonia Fahmy, Raj Jain, Shivkumar Kalyanaraman, Rohit Goyal, Bobby Vandalore and
Xiangrong Cai
The Ohio State University
Department of Computer and Information Science
Columbus, OH 43210-1277
Phone: 614-292-3989, Fax: 614-292-2911
Email: {*fahmy, jain, shivkuma, goyal, vandalor, cai*}@cis.ohio-state.edu

Seong-Cheol Kim
Principal Engineer, Network Research Group
Communication Systems R&D Center
Samsung Electronics Co. Ltd., Chung-Ang Newspaper Bldg.
8-2, Karak-Dong, Songpa-Ku, Seoul, Korea 138-160
Email: kimsc@metro.telecom.samsung.co.kr

**Abstract-** Asynchronous transfer mode (ATM) networks must define multicast capabilities in order to efficiently support numerous applications, such as LAN emulation, Internet protocol (IP) multicasting, video conferencing and distributed applications. Several problems and issues arise in ATM multicasting, such as signaling, routing, connection admission control, and traffic management problems. IP integrated services over ATM poses further challenges to ATM multicasting. Scalability and simplicity are the two main concerns for ATM multicasting. This paper provides a survey of the current work on multicasting problems in general, and ATM multicasting in particular. A number of proposed schemes is examined, such as the schemes MARS, MCS, RSVP, SEAM, SMART, and various multipoint traffic management schemes. The paper also indicates a number of key open issues that remain unresolved.

---

[1] May 15, 1997. Under preparation. Available through http://www.cis.ohio-state.edu/~jain/papers/mcast.htm

# Contents

# 1   Introduction

Asynchronous Transfer Mode (ATM) is the technology of choice for the Broadband Integrated Services Digital Network (B-ISDN). ATM is proposed to transport a wide variety of services in a seamless manner. In ATM, user information is transferred in a connection oriented fashion between communicating entities using fixed-size packets, known as ATM cells. The ATM cell is fifty-three bytes long, consisting of a five byte header and a forty-eight byte information field, referred to as the payload.

A truly efficient, flexible and scalable ATM multipoint service is a key factor in the success of ATM networks. ATM multicasting is essential for several applications, such as LAN Emulation (LANE) and IP multicasting over ATM applications, in addition to future audio and video conferencing and video distribution applications. Defining ATM multicasting is a challenging task. Several issues need to be addressed, such as routing, signaling, resource reservation, traffic management and providing reliable transport. Providing IP multicast over ATM poses further problems. This paper surveys the work that has been done in multicasting, and points out a number of issues that need to be more carefully investigated.

The remainder of this paper is organized as follows. The next section discusses IP multicasting, and various proposals for IP multicasting to operate over ATM. Then, several signaling issues are discussed, and in particular, the cell interleaving problem is examined. ATM traffic management for multipoint connections is then explored in detail, and a number of proposals for modifying ATM switch schemes are presented. The ABR source rule parameters and the performance of the schemes are briefly highlighted. Issues pertaining to real-time multipoint traffic and resource reservation are also explored, and the future work in that area is discussed. Transport protocol proposals for multipoint traffic are then compared, and interoperability issues are briefly mentioned. The paper concludes with a discussion of the open issues in ATM multicasting and a summary of the survey.

# 2   Overview of IP Multicasting and ATM Multicasting

Multicasting in the Internet Protocol (IP) has been defined in 1989 by specifying the extensions that the IP host needs to implement, as well as the behavior of the multicast routers. ATM multicasting, on the other hand, is still in earlier phases of definition. Supporting IP multicasting over ATM has been the subject of extensive research, since the currently defined ATM User Network Interface (UNI) provides limited multipoint capabilities. This section outlines IP multicasting, and then proceeds to examine a number of proposals for supporting IP multicasting over ATM.

## 2.1 IP Multicast

IP multicast is based on the Internet Group Management Protocol (IGMP), and routing is commonly implemented by one of the Internet multicast routing protocols, such as the Distance Vector Multicast Routing Protocol (DVMRP). This subsection briefly overviews IP multicasting.

Internet host protocols have been extended to support IP multicast. IGMP allows IP hosts to join and leave multicast groups. The membership of multicast groups is dynamic, and there is no restriction on the location or number of members in a host group. A host can be a member of any number of groups, and need not be a group member to send datagrams to a group. Host groups can be permanent or transient. Permanent groups can have any number of members at any time, even zero, but transient groups only exist as long as they have members. "Multicast routers" ("mroute" routers) handle the forwarding of datagrams and propagation of routing information. Three levels of IP multicast conformance are defined: no support, send and not receive, and full support.

Multicast IP addresses start with the reserved 4-bit sequence 1110 (class D IP addresses), and the rest of the address (the remaining $32 - 4 = 28$ bits) indicates the multicast group number. Group 1 denotes the permanent group of all hosts on this net. Several extensions to IP, the IP interface, and the network interface are implemented to support IP multicast. The underlying Ethernet (or local net) multicast is used, and IP multicast addresses map to the Ethernet multicast address space. The routines "JoinHostGroup" and "LeaveHostGroup" are specified at both the IP and the network interfaces. IGMP provides messages used to query hosts about their group memberships. Only one host per net need reply. The queries are periodically broadcast to the net. A random timer is used to prevent collisions. Hosts only need to inform routers of join requests, and not leave requests [15].

Many-to-many IP multicast can use two approaches for data distribution, namely: the shared tree approach and the (per) source-based tree approach. The shared tree approach uses a common multicast tree that is shared by all sources (senders), whereas the source-based tree approach requires each source to maintain its own multicast tree. Core-Based-Trees (CBTs), and Protocol-Independent-Multicast (PIM)-Sparse Mode (SM) are examples of the shared tree approach, while the Distance Vector based Multicast Routing Protocol (DVMRP), Multicast extensions to Open Shortest Path First (MOSPF), and Protocol-Independent-Multicast (PIM)-Dense Mode (DM) PIM are examples of the source-based tree approach. There are several ways in which the multicast trees can be created, namely "broadcast-and-prune," "link-state broadcast of receiver joins" and "explicit join by the receivers" [11].

The Core-Based-Tree (which is a shared tree idea) is one of the most popular approaches. This is because it is not too difficult to implement. The non-optimality of routing in this approach is not a major issue when there is a large number of flows present. In addition, this approach represents a fairly simple mechanism of managing multicast data distribution trees [11]. Refer to [9] for more information on the requirements for multicast protocols, including routing protocols, and group address and membership authority [9].

## 2.2 IP Multicast over ATM

IP over ATM has been extensively discussed at the Internet Engineering Task Force [33, 34], as well as at the ATM Forum multiprotocol over ATM group [19]. RFC 1932 [13] compares all the IP over ATM proposals. IP multicasting over ATM poses a set of new challenges to the IP over ATM problem, especially with the limited ATM multipoint capabilities defined in UNI 3.0/3.1 and even UNI 4.0. A number of proposals have addressed these problems [19]. Signaling support for IP over ATM has been extensively discussed in [38, 36] and is further discussed in the next section.

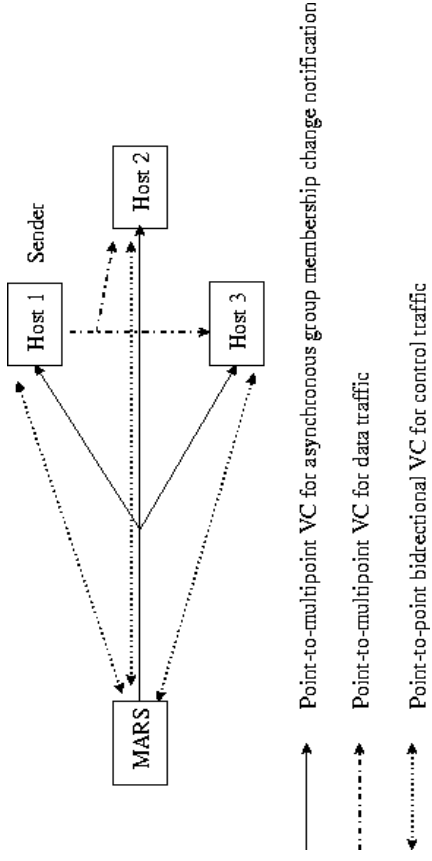

Figure 1: The MARS architecture

ATM-based IP hosts and routers can use a multicast address resolution server (MARS) to support IP multicast over ATM UNI 3.0/3.1 point-to-multipoint connection service. The MARS approach builds on the classical model of IP over ATM, emulating a shared media broadcast LAN over a non-broadcast ATM network. The MARS manages a cluster of ATM end points, where a cluster is defined as the set of ATM interfaces choosing to participate in direct ATM connections to achieve multicasting among themselves [1].

Clusters of end points can use MARS servers to track receiver nodes for given multicast groups, as illustrated in figure 1. The MARS maps the layer 3 multicast addresses into the ATM addresses of the corresponding group members. Thus, end points can establish and manage point-to-multipoint Virtual Circuits (VCs) when transmitting to the group. Each host in a multicast group also establishes a point-to-point bidirectional VC to the MARS for control traffic, and the MARS asynchronously notifies hosts of group membership changes using a point-to-multipoint VC called the cluster control VC (figure 1) [1, 2].

Proponents of the MARS approach argue that the MARS model simplifies the system by dividing it into clusters connected by mrouters, and using mrouters to forward packets outside a cluster. The key point when overlaying the classical IP routing model on top of ATM clouds, is that the mrouters aggregate the data flow and hide the group membership information. Short-cutting the multicast link-level distribution paths (by bypassing mrouters) completely eliminates these two unique characteristics. The MARS model only needs to track intra-cluster memberships, and handles the problems with interleaving cells from different packets at the ATM adaptation layer. This allows traffic from one cluster to be aggregated and forwarded on a single VC to reach another cluster. Bypassing the mrouter loses the aggregation point, which leads to VC explosion and a large VC mesh. Another problem with short-cutting mrouters between clusters is the tracking of membership state [11].

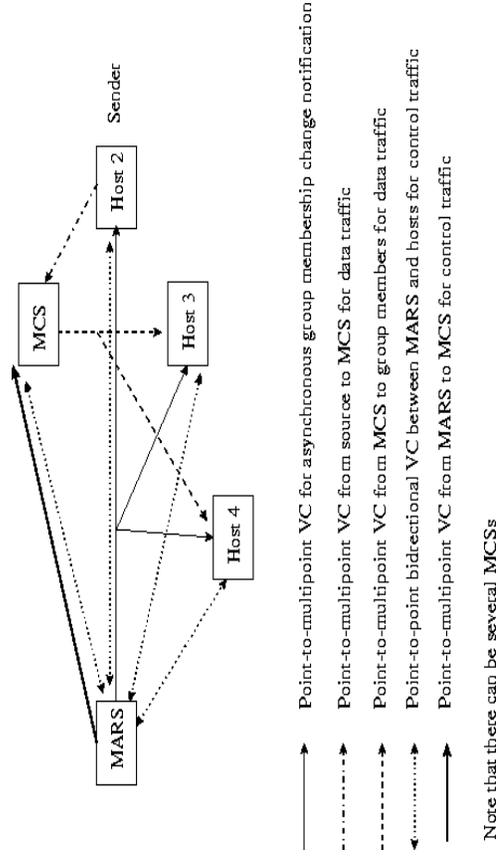

Figure 2: The MCS architecture

For a sender to transmit data to a group of receivers, it can set up a point-to-multipoint VC that connects it to all the multicast group reveivers (as in figure 1). This approach is called a "VC mesh", because each of the senders in the group usually establishes a VC to all the group members, requiring a mesh of VCs for the group to communicate. An alternative approach to this uses a multicast server (MCS) as shown in figure 2. The MCSs can receive data traffic from senders and multicast it to the receivers of the multicast. A point-to-multipoint VC must be set up from the MARS to the MCSs for control traffic.

However, the multicast server can be a bottleneck, so multiple multicast servers may be required in the same multicast group. The senders and receivers are oblivious to the existence of multiple MCSs. Multiple MCSs make the MCS architecture more fault tolerant, and prevent bottlenecks by load sharing. In addition, multiple MCSs minimize the changes needed to MARS clients.

To synchronize among multiple MCSs, the MCSs can employ a synchronization protocol, whereby when one MCS is active, the others act as backups. Alternatively, the MARS can distribute senders/receivers on a per-group basis, since it has all the necessary information in a consistent state. In this case, the MARS maintains an allocation map that keeps track of group distribution, and uses a point-to-point VC to every MCS to communicate group membership changes. Using the MARS for synchronizing MCSs has the advantages of ensuring consistency and avoiding the need for a separate allocation entity [52].

The main advantage of the MCS approach over the VC mesh approach is the reduction in the number of VCs. Even if the number of senders is small and group membership is small, the signaling load on the network for a VC mesh can be significant. Simulations results indicate that the mesh needs significantly more VCs than the MCS: more aggregate VCs in the network, and more per-host VCs [52]. A comparison of the two IP multicast over ATM approaches is provided in [10]. The next section surveys some of the native ATM multicast solutions to such problems.

# 3 ATM Multipoint Definition: Signaling, Forwarding and Routing Issues

At this time, UNI signaling supports multicast via point-to-multipoint VCs. The ATM UNI 3.1 signaling standard supports the source-based tree approach of multicast data distribution and uses root-initiated joins to multicast tree construction. Since receiver or leaf initiated join (LIJ) is a more scalable approach, UNI 4.0 signaling supports such joins [47]. However, a pure multipoint-to-multipoint service is not yet supported.

Furthermore, the ATM private network to network interface (PNNI) 1.0 does not define routing for multipoint connections. The second phase of PNNI will define routing for UNI 4.0 multipoint connections [20, 18].

As previously discussed, the MARS architecture uses the point-to-multipoint approach. It uses the point-to-point and point-to-multipoint VCs supported by UNI 3.1 signaling to forward packets within a cluster, and uses a multicast router to go outside a cluster. However, in this scenario, the sources or servers need to know which receivers are listening to which multicast group. This incurs a state overhead as well as a state management overhead, leading to scalability problems for very large multicast groups. Another problem is the ability of the receiver in a multicast group to distinguish cells coming from different concurrent senders [11].

A number of proposals have attempted to attack some of the problems that the MARS and MCS proposals have attacked without requiring the use of a dedicated server. Such proposals attempt to provide a scalable architecture for ATM multipoint-to-multipoint, without a special server to handle forwarding, while avoiding the scalability problems of VC meshes.

One of the main problems to be solved is the cell interleaving problem. Solutions to the cell interleaving problem attempt to prevent interleaving of cells of packets originating from different sources on the same multipoint connection. As shown in figure 3, the cells of a packet should not be interleaved with cells from another packet from a different sender after merging. Since ATM adaptation layer 5 (AAL5), which is most commonly used with data traffic does not contain any multiplexing identifier or sequence number, and all traffic within the group uses the same VC identifier, alternate solutions must be implemented.

Potential solutions to the cell interleaving problem with AAL5 include: [11, 54, 53]:

1. Overlay one-to-many VCs to create many-to-many multicast (VC mesh): This solution

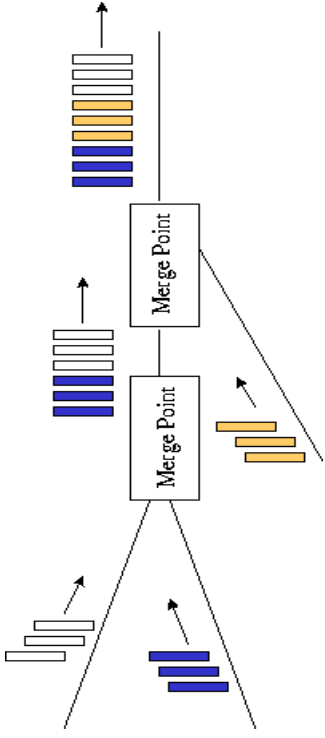

Figure 3: The cell interleaving problem

does not scale and requires N one-to-many VCs for N senders.

2. Use control messages to coordinate senders, as in [23, 24]: This has a high overhead.

3. Use AAL3/4: This suffers from excessive overhead and is not well supported.

4. Multipoint Virtual Paths (VPs): This requires receivers to have static assignment of VCs within VPs. In addition, VPs should not be used by end-systems, as network providers use them for aggregation in the backbone.

5. Packet-level buffering, as in [25]: This is achieved by buffering cells of other VCs till all cells of this VC go through. The technique is called "cut-through forwarding" [25]. It entails the implementation of a packet-based scheduling algorithm at the merging point, and maintaining separate queues for each sender. The AAL5 end of packet cell is used to signal to the switch that a packet from a different port can be forwarded. However, this requires more memory at the switches, and adds to the burstiness and latency of traffic.

6. Dynamic sharing by using on-the-fly mapping of a sub-channel [11]: A sub-channel is a "channel within a VC." Each sub-channel is assigned an identifier called the sub-channel number to distinguish between multiple sub-channels in a VC. Four bits from the Generic Flow Control (GFC) bits in the ATM cell header can carry this number. However, four bits allow only up to sixteen concurrent senders. Hence, this solution is clearly not scalable.

It is clear that each of these methods has its own merits and drawbacks.

Another problem in ATM multipoint-to-multipoint communication is the routing problem. It is important to allow member-initiated joins of the multicast group. Some schemes, such as the SEAM scheme [25] use a single shared tree for all senders and receivers, and use a CBT-like core for routing (see figure 4). This idea is borrowed from IP. In SEAM, a single VC is used for a multicast group consisting of multiple senders and multiple receivers. Each conversation in SEAM has a unique identifier called a group handle, which identifies all packets associated with a given multicast group. The core is used as a focal point for routing signaling messages for the group. Both member-initiated and core-initiated joins of the multicast group are allowed. Little changes in signaling are required. A technique called "short-cutting" [25] is also used whereby cells are forwarded by switches on a spanning tree. This mechanism allows packets to follow the shortest path along the shared tree by emulating reverse path forwarding.

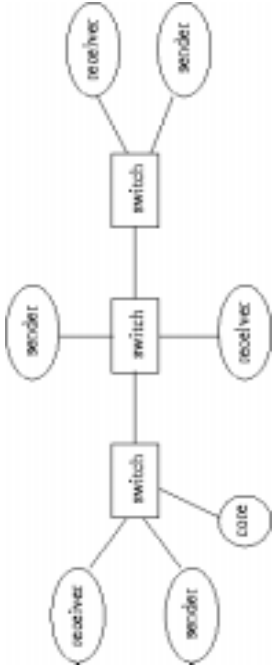

Figure 4: The core in a multipoint connection

The Shared Many-to-Many ATM Reservations (SMART) [23, 24] protocol is also scalable, like SEAM, in the sense that many-to-many VC connections are used. Thus the number of VC connections is independent of the number of endpoints. The SMART scheme can interoperate with non-SMART switches. SMART serializes the communication of various senders to a set of receivers, completely avoiding the cell interleaving problem. Since SMART defines a number of control messages (implemented as special types of ATM Resource Management cells) that guarantee the serialization of communication, it not only solves the cell interleaving problem, but also guarantees that the traffic contract associated with the VC connections is respected, thus performing a very simple traffic management function. The next two sections discuss ATM traffic management in further detail.

# 4 ATM Traffic Management for Multipoint Connections

ATM networks offer five classes of service: constant bit rate (CBR), real-time variable bit rate (rt-VBR), non-real time variable bit rate (nrt-VBR), available bit rate (ABR), and unspecified bit rate (UBR). Of these, ABR and UBR are designed for data traffic, which exhibits a bursty unpredictable behavior.

The UBR service is simple in the sense that users only negotiate their peak cell rates (PCR) when setting up the connection. If many sources send traffic at the same time, the total traffic at a switch may exceed the output capacity causing delays, buffer overflows, and loss. The network tries to minimize the delay and loss but makes no guarantees.

The ABR service provides better service for data traffic by periodically advising sources about the rate at which they should be transmitting. The switches monitor their load, compute the available bandwidth, and divide it fairly among the active virtual connections (VCs). The feedback from the switches to the sources is sent in Resource Management (RM) cells which are sent periodically by the sources and turned around by the destinations (see figure 5).

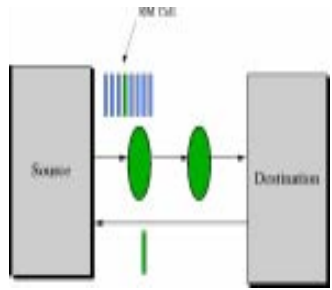

Figure 5: RM cell path

The RM cells contain the current cell rate (CCR) of the source, in addition to several fields that can be used by the switches to provide feedback to the sources. One of those fields, the explicit rate (ER) field, indicates the rate which the network can support at that time. Each switch on the path of the VC reduces the ER field to the maximum rate it can support. The sources examine the returning RM cells and adjust their transmission rates accordingly.

The RM cells flowing from the source to the destination are called forward RM cells (FRMs) while those returning from the destination to the source are called backward RM cells (BRMs). When a source receives a BRM, it computes its allowed cell rate (ACR) using its current ACR, congestion indication flags in the RM cell (congestion indication, CI, and no increase, NI), and the explicit rate (ER) field of the RM cell. The ER field indicates the rate that the network can support at the particular instant in time.

The ATM Forum traffic management specification currently provides some guidelines on traffic management of point-to-multipoint connections, but does not enforce nor suggest a specific strategy. Congestion control strategies for multipoint-to-point, and multipoint-to-multipoint connections are still under study [16].

## 4.1 Point-to-Multipoint Connections

The traffic management problem for point-to-multipoint connections is an extension to the traffic management for unicast connections. However, some additional problems arise in the point-to-multipoint case. In particular, the consolidation of the feedback information from the different leaves of the tree is necessary for point-to-multipoint connections (see figure 6). This is because of the "feedback implosion" problem (feedback information provided to the sender should not increase proportional to the number of leaves in the connection). Scalability becomes a major concern.

Many general frameworks have been suggested that convert any unicast congestion control switch algorithm to work for point-to-multipoint connections [44, 49, 40]. In addition, several issues pertaining to source end system parameters for point-to-multipoint connections have been discussed [46, 29]. This section highlights the major issues in this area.

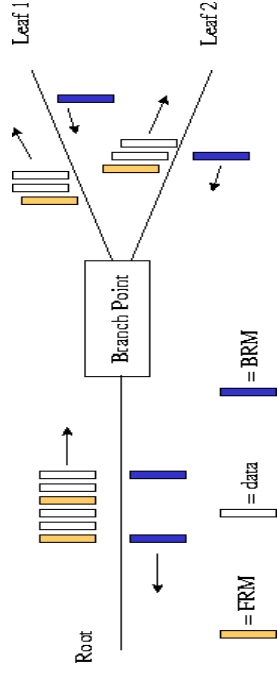

Figure 6: Point-to-multipoint connections

### 4.1.1 Source Parameters

Setting ABR source parameters for point-to-multipoint connections was briefly examined in [46], [7] and [29], but the results were not incorporated in the ATM traffic management specifications [21] because more studies need to be conducted to determine the best method and guidelines to compute their values. The main factor that complicates the setting of these parameters for multipoint connections is the possible existence of widely varying round trip times from the source to the different receivers, and the possible existence of a bottleneck on a distant branch. Thus a more conservative approach in the setting of the parameters might be preferred. However, this can adversely affect the efficiency experienced by the connections.

Multipoint connections may suffer from initial overallocation until feedback is received from all the distant leaves. Initial overallocation can be overcome by correct setting of the cells in flight (CIF) parameter and the correct calculation of the initial cell rate (ICR) parameter. In [46], a formula is proposed to calculate the optimal value of ICR for multipoint connections, and an approximation is proved to achieve an approximately equivalent performance. In this formula, the value of ICR is a function of the CIF value, the longest round trip time (RTT) value, and the rate increase factor (RIF).

We note that if the calculation of ICR depends on the round trip time, a problem arises: should ICR change when nodes join or leave the group to account for the longest RTT for all destinations? The RTT of the farthest leaf reduces the ICR value according to the proposed formula. What happens when that farthest leaf leaves the multicast group? The root should not be notified every time a leaf joins or leaves the group, otherwise the signaling rules would imply that ABR multicast will not scale [29].

In addition, most of the previous studies [46] argue that transient queues can be mitigated by setting the RIF ABR source parameter to a small value. This can be especially useful in cases of distant bottlenecks that are multiple branch points away. Although conservative values are advisable for multipoint connections (where feedback response from distant bottlenecks is not always available), such small values have the adverse effect of slowing the rise to the optimal value. Such tradeoffs need to be further investigated.

### 4.1.2 Consolidation of Feedback Information

One of the common goals for point-to-multipoint ABR is to ensure that all destinations receive all cells from the source. This requires that the source be controlled to the minimum rate supported by all the destinations. The minimum rate is the technique most compatible with the typical data requirements where no data should be lost and the network can take whatever time it requires to deliver the data intact. This is especially useful for LANE and non-real-time data services [45].

Hence, the source in a point-to-multipoint VC should send at the minimum of the rates allowed by all the destination nodes. The first proposal to control the source rate in this manner can be found in [44]. The method works as follows. A register MER is set to min(MER,ER in BRM cell) whenever a BRM cell is received from one of the branches. When an FRM cell is received, it is multicast to all branches, and a BRM is sent with the MER value as the ER indicated by the network. MER is then reset to the ER value in the FRM cell (typically PCR). Thus the minimum of all the rates supported by any branch is selected and returned to the source.

At the ATM forum, the consolidation algorithm was proposed to be implementation-specific with the previously explained algorithm given as an example consolidation. The source and destination behaviors are unaltered, and the consolidation algorithm is optional. Sources may also need to do the consolidation in that case [3].

In [49], the multipoint extension was applied to an ABR rate allocation algorithm, and the results show that the extended algorithm is max-min fair if the point-to-point algorithm is max-min fair. Necessary and sufficient conditions are defined for max-min fairness in point-to-multipoint cases, which is an intuitive extension of the unicast max-min definition. A proof is provided that shows that the previously discussed algorithm for deriving a multipoint congestion control algorithm from a unicast one preserves the efficiency and fairness properties of the unicast algorithm.

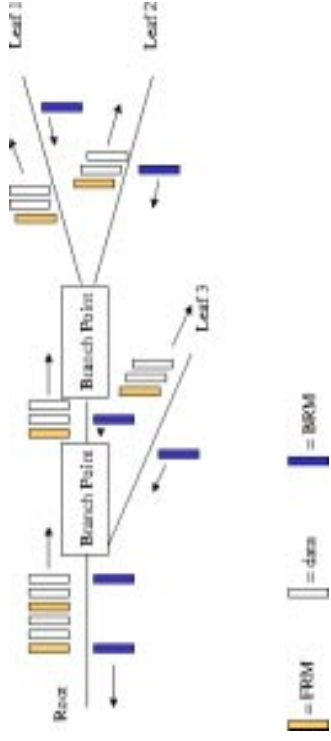

Figure 7: Multiple branch points

It is important to study the effect of multiple branching points. Figure 7 illustrates a multicast connection with two branching points. Observe that at each switch node, an additional cycle of N cells is required in order to accumulate the information from the branches. Thus if a multicast tree has 4 levels of branching, then the information from the lowest branches will take 4N cell times to propagate back to the source (as opposed to N cell times in the point-to-point case).

As a result of this additional delay, the responsiveness of the multicast algorithm will be worse than that for the point-to-point VCs. Hence, buffer allocations for the multicast queues will have to be somewhat higher, since it takes longer for congestion information from one branch to reach the source [44, 7].

Another (more serious) problem may also arise. The point-to-multipoint framework generates a BRM cell from a branch point to the root when an FRM arrives, and the BRM contains the consolidated information from the branches that provided feedback after the last BRM was sent. As a result, the BRM, in general, does not capture feedback information from all branches. This introduces noise called *"consolidation noise,"* and is mainly caused by the asynchrony of feedback and the rate fluctuations.

Due to this, Hunt [29] argues that an existence proof is necessary for a multicast mechanism suitable for the expected range of traffic patterns, number of VCs, bandwidth bottlenecks, and round trip times. Bursty traffic sources, as well as a wide potential range of RTTs from the source to the various leaves (RTT difference should be a couple of orders of magnitude) should be examined, but a worst case analysis needs to be provided. Cases to be tested include (1) bursty traffic models (2) dynamic CBR and VBR in the background on bottleneck links (3) many ABR VCs, especially point-to-multipoint VCs (4) several orders of magnitude variation in the available bandwidth at the bottlenecks (5) changes in dynamic capacity at transient periods, as well as in steady state, and (6) a large RTT ratio from the nearest to the farthest leaf [29]. Other problem situations should also be analyzed, and metrics to use should be developed.

Several variations on the Roberts algorithm [44] have been proposed [40]. They employ other approaches to consolidate the feedback information in the multicast tree. These algorithms mainly differ in whether the switch needs to generate BRM cells or not, and in whether the switch should wait for feedback from all the leaves of the multicast tree before sending feedback to the source. Some of the new schemes are simpler to implement than the previous proposal that required the branch point to generate a returning RM cell for every forward RM cell, while others achieve better performance. These variations are examined in the following few paragraphs.

The first modification tries to alleviate the "consolidation noise" problem. As previously mentioned, the early proposal suffers from consolidation noise, where a BRM generated by switch may not consolidate feedback from all tree branches. In fact, if a BRM generated by a switch does not accumulate feedback from any branch, the feedback can erroneously be given as the peak cell rate, or the ER supported at this branch point (which may be very high). A simple enhancement to avoid this problem is to maintain a flag, and only generate the BRM cell if a BRM has been received from a leaf since the last BRM was sent by the branch point [40].

Another idea reduces the complexity of the algorithm as follows. The backward RM cells are generated solely by the destinations and NOT by the switches, which is similar to the case of unicast [39]. The motivation behind this modification is as follows. If switches turn around RM cells, the implementation has a high cost and complexity. In the earlier algorithms, the number of BRMs generated by switches per forward RM cell from the source is proportional to the number of branch points in a multicast tree. The new algorithm proposed that does not generate BRMs at branch points whenever FRMs are received, but simply sets a flag indicating the receipt of the FRM and broadcasts it to all leaves. When a BRM is received from a branch, it is passed back to the source (after using the minimum allocation), only if the previously mentioned flag was set. The flag is then reset, as well as the MER value [40].

It is natural to extend this idea to only send back the BRM when BRMs from *all* branches are received. This can be easily implemented by maintaining a separate bit for each branch that indicates if a BRM has been received since the last BRM was sent. Clearly this method incurs additional complexity, compared to the previous one. Moreover, it has to deal the problem of failure of one of the branches by implementing timeouts. The four variations of the algorithm were compared in [40] While consolidation noise was least with the last method, the additional complexity might not be worth the benefits, especially that the method exhibits a slow transient response.

A similar method to the latter method was proposed in [12]. Again, the algorithm only allows feedback to return to the source when BRMs have been received from all branches. However, the scheme proposes to add a sequence number to the RM cells. The BRM that is allowed to pass back to the source is the last BRM to be received with a certain sequence number. This guarantees that among all BRM cells with the same sequence number, one and only one BRM passes back to the source, and that BRM is the BRM of the destination with the longest RTT. This is independent of the number of branch points in the tree. The returning BRM collects the latest feedback indicated by all branches. Clearly, this method is even more complex than the one proposed in [40], and suffers from an initially slow transient response.

In conclusion, the different variations exhibit a tradeoff between complexity, transient response and minimization of consolidation noise. We believe that a number of other issues should also be studied, such as the effect of multiple branching points, and the tradeoffs should be studied under a large variety of conditions to determine the best approach.

## 4.2 Multipoint-to-Point Connections

Little work has been done to define traffic management rules for multipoint-to-point connections [28]. Because the traffic at the root is the sum of all traffic originating at the leaves, bandwidth management is an important issue. An important issue in the case of multi-ple senders is how to define max-min fairness within a multicast group and among multicast groups and point-to-point connections. Billing and pricing issues may play an important role in such cases. Bear in mind that the multicast connection has the same identifier (VPI/VCI) on each link and sources cannot be distinguished, yet allocation should be max-min fair.

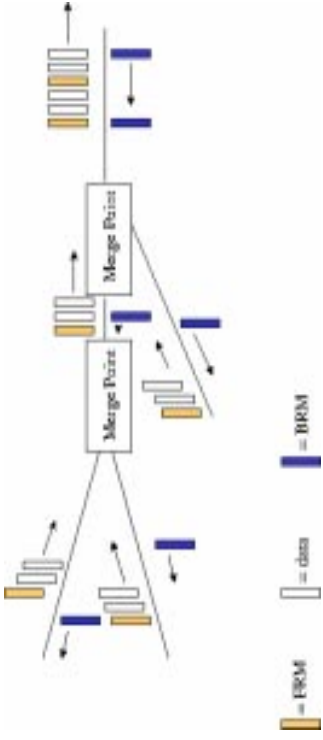

Figure 8: Multipoint-to-point connections

As illustrated in figure 8, multipoint-to-point connections require feedback to be returned to the appropriate sources at the appropriate times. Note that the bandwidth requirements for a VC after a merge point should be the sum of the bandwidths used by all senders whose traffic is merged [31].

Ren describes an algorithm for multipoint-to-point congestion control, which allows heterogeneous senders belonging to the same connection with different data rates. The author proved that if the original point-to-point switch algorithm is max-min fair, the multipoint-to-point version is also max-min fair [39, 41, 42]. The idea of the algorithm is very similar to the point-to-multipoint algorithm previously discussed. An FRM cell is processed normally and forwarded to the root, also returning a BRM cell to the source which sent the FRM cell. The receipt of a BRM at the splitting point simply results in the normal processing of the BRM, and using the values it contains to set the value in the MER register. The BRM cell is then discarded.

Another alternative would be to maintain a bit at the merge point for each of the senders. The bit indicates that an FRM has been received from this sender after a BRM was sent to them. Therefore, when an FRM is received, it is forwarded to the root and the bit is set to 1. When a BRM is received, it is duplicated and sent to the branches that have their bit set, and then the bit is reset. This saves the overhead that the switch incurs when it turns around RM cells [43].

Note that both the algorithms should preserve the fairness and efficiency properties of the original point-to-point algorithm, and do not assume knowledge of the number of senders. The aggregate data rate after a merging point is the sum of all incoming data rates to the merging point. Similarly, the number of RM cells after merging is the sum of those from different branches. Hence, the ratio of RM to data cells remains the same. A number of simulation results illustrate that the scheme adequately works for peer to peer and parking lot configurations with an infinite traffic pattern [41, 42, 43]. More extensive analysis needs to be performed for this and similar frameworks, and the transient performance and delays need to be further examined.

## 4.3 Multipoint-to-Multipoint Connections

The development of traffic management schemes for multipoint-to-multipoint connections (figure 9) is still under study. Extensive performance analysis must be carried out to ensure that congestion can be avoided for many-to-many VCs, regardless of the topology of the senders and receivers in the connection, their round trip times and bottleneck locations, the traffic characteristics and the background traffic patterns.

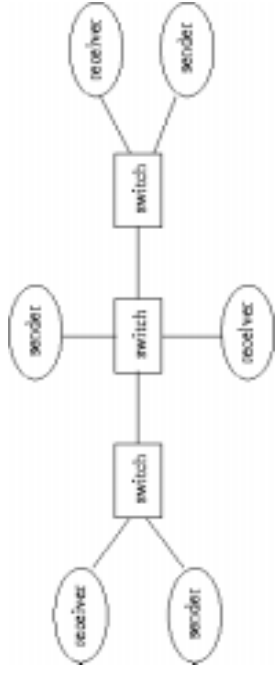

Figure 9: Multipoint-to-multipoint connections

Simple schemes like SMART [23, 24] use control cells and state information at the nodes of the multicast tree to ensure that the traffic contract is respected. The main idea of the SMART scheme is that the resources for the requested service are reserved, and RM cells are used as special control messages (called "GRANT" and "REQUEST" messages) to control the link access. These message are associated with each VC connection. When a system receives a "GRANT" message, this means that the sender of the "GRANT" is willing to receive data on this VC connection. If two "GRANT" messages coming from different directions cross each other on a link, a "bias" (initially negotiated among every two neighbors) is used to resolve the conflict. In addition, RM cell races are avoided by using a two-bit sequence number in the RM cells. The protocol grants requests in such a way that the reservations are respected, and the cell interleaving problem previously described can never occur. The cell rates are determined by the protocol, based upon the requests by the users and the decisions taken by intermediate systems.

Schemes such as SMART might exhibit a high overhead resulting from the large number of control messages exchanged before sending data. In [43], ABR multipoint-to-multipoint congestion control is developed as a combination of [40] and [42] as explained in the previous two subsections. The connection is assumed to be set up as a shared tree and merging/branching points combine the algorithms presented in the last two subsections. It seems reasonable for multipoint ABR flow control to be implemented by combining good point-to-multipoint and multipoint-to-point algorithms, so that the point-to-point algorithm would simply be a special case of the multipoint one.

## 4.4 UBR Performance

Preliminary results of UBR performance for multipoint-to-point connections can be found in [41]. The multipoint-to-point connections in peer to peer and parking lot configurations were simulated. The preliminary results indicate no problems, except when scheduling packets of different sizes from different senders with cut-through forwarding. In that case, unfairness can be seen, which is to be expected when packet round-robin is used with different packet sizes.

# 5 Quality of Service (QoS) Requirements for Multipoint Connections

As discussed in the previous section, traffic management for multipoint connections can be an extremely challenging problem. A further challenge to traffic management is introduced by real-time traffic. This section discusses resource allocation for multipoint connections, highlighting the resource reservation protocol proposed for IP.

## 5.1 Heterogeneity and Dynamic Behavior

In IP, different receivers in a multicast group can specify different quality of service (QoS) requirements. When a virtual connection includes receivers with different QoS requirements, the VC is commonly referred to as a "variegated VC." In addition, receivers are allowed to dynamically change their QoS requirements throughout the connection lifetime. Group membership also changes throughout connection lifetime.

ATM does not currently allow different destinations in a multicast group to have different QoS requirements, or different senders to specify different traffic characteristics. Renegotiation is also not currently allowed. However, variegated VCs are important for various ATM classes of service (at least VBR and ABR), at least because IP allows such heterogeneity, and mapping techniques that map such variegated IP traffic to ATM are not entirely flexible and introduce heavy overhead (see [4] and [22]). Dynamic behavior is currently foreseen to be supported in ATM by tearing down the ATM connection and setting up a new one, which is clearly inefficient. ATM must adapt to the dynamic and varying needs of receivers in an IP multicast connection, and it must directly support those needs [14, 31, 16, 6].

Techniques such as hierarchical encoding, translators and intelligent drop policies can dynamically provide different receivers with different QoS. With video traffic, using techniques such as translators, interlacing (used in GIF), progressive and hierarchical encoding (used in JPEG and MPEG), and intelligent scheduling and drop policies can be used to produce data at different rates to different receivers in the same multicast group. For example, the receiver that can only receive at the slowest rate can receive only the highest priority traffic, while the receiver that can receive at the highest rate can receive all levels of the encoded traffic.

The signaling rules of ATM can be modified to allow renegotiation of parameters. Connection admission control procedures can be streamlined to enable this to be accomplished with minimum overhead. One way to do this might be to simply propagate the resource reservation requests during the connection lifetime and perform a subset of the connection admission control functions.

## 5.2 Resource Reservation Protocol (RSVP)

The resource reservation protocol (RSVP) has been developed to support traffic requiring a guaranteed quality of service over IP multicast. RSVP can operate transparently through routers that do not support it. This is because RSVP is compatible with existing network infrastructures.

The guaranteed quality of service requirements are detailed in [48]. RSVP interacts with the packet schedulers at the routers to ensure that QoS requirements are met [30, 4, 55]. The RSVP protocol provides receiver-initiated setup of resource reservations for both unicast and multicast data flows. RSVP operates on top of IP, and is only concerned with the QoS of the packets forwarded according to routing.

RSVP interacts with a packet classifier and a packet scheduler to determine the route and achieve the required QoS. An RSVP reservation request consists of a "flowspec", specifying the desired QoS, as well as a "filterspec", defining the flow to receive the desired QoS. The packet scheduler is completely responsible for negotiation [8].

RSVP uses "soft state," and sends periodic "refresh" messages to maintain the state along the reserved paths. Thus, it can adapt dynamically to changing group membership and changing routes.

RSVP is simplex (unidirectional), and supports several reservation styles to fit a variety of applications. Reservation styles supported by RSVP include wildcard filters, which select all senders, reserving resources to satisfy the largest resource request, regardless of the number of senders. Another type of reservation style is the fixed filter, which creates one reservation per specified sender, without installing separate reservations for each receiver to the same sender. The last type of reservation style is the dynamic filter, where each reservation request can specify several distinct reservations to be made using the same flow specification. The number of actual reservations made in this case depends on the number of senders upstream [50].

RSVP receivers use the reserve (RESV) message to periodically advertise to the network their interest in a flow, specifying the flow and filter specifications. RSVP senders, on the other hand, send a PATH message to indicate that they are senders, and give information such as multicast address, reservation identifier (ID), previous hop IP address, templates for identifying traffic from that sender, and flow specification. The message is sent to all receivers in the multicast tree. The network is free to accept the reservation, reject it, or reduce the requirements [50].

RSVP is compatible with existing network infrastructures. To guarantee the bandwidth and delay characteristics reserved by RSVP, a fair scheduling scheme, such as Weighted Fair Queuing (WFQ) can be employed. WFQ isolates data streams and gives each a percentage of the bandwidth on a link. This percentage can be varied by applying weights derived from RSVP's reservations. As previously mentioned, applications that receive real-time traffic inform networks of their needs, while applications that send real-time traffic inform these receivers about the traffic characteristics they may specify.

To summarize, RSVP receivers periodically alert networks to their interest in a data flow, using RESV messages that contain the source IP address of the requester and the destination IP address, usually coupled with flow details. The network allocates the needed bandwidth and defines priorities. Eventually, an RSVP receiver stops advertising its interest in a flow. An RSVP sender uses the PATH message to communicate with receivers informing them of flow characteristics. The "soft-state" feature allows networks to be self-correcting despite routing changes and loss of service. This enables routers to understand their current topologies and interfaces, as well as the amount of network bandwidth currently supported. RSVP-equipped routers can adjust network capacity in real time [51, 50]. The next subsection discusses how RSVP can make use of ATM quality of service.

## 5.3 RSVP over ATM

Since ATM networks provide QoS guarantees, it is natural to map RSVP QoS specifications to ATM QoS specifications, and establish the appropriate ATM switched virtual circuits (SVCs) to support the RSVP requirements. However, the problem is complicated by several factors that were mentioned before: RSVP allows heterogeneous receivers and reservation parameter renegotiation, while ATM does not. The solution for providing RSVP over ATM must tackle these problems, ensuring scalability. It must also support both UNI 3.1 and UNI 4.0, which only support point-to-multipoint connections [14, 5, 4, 36, 22].

The problem of supporting RSVP over ATM consists of two main subproblems: first, mapping the IP integrated services to ATM services, and second, using ATM VCs with QoS as part of the integrated services Internet [22, 4, 5].

The mapping of IP integrated services to ATM services is explained in [22]. It is not a straightforward task, and has many facets. The IP services considered are guaranteed service, and controlled load service, in addition to the default best effort service. The guaranteed service is mapped to CBR or VBR-rt, the controlled load service is mapped to VBR-nrt or ABR with a minimum cell rate, and the best effort service is mapped to UBR or ABR. A number of parameter mappings and signaling element mappings are needed for service interoperation.

The second subproblem, managing ATM VCs with QoS as part of the integrated services Internet, entails deciding upon the number of VCs needed and designating the traffic flows that are routed over each VC. Two types of VCs are required: data VCs that handle the actual data traffic, and control VCs which handle the RSVP signaling traffic [4]. The control messages can be carried on the data VCs or on separate VCs.

As previously mentioned, it is essential to tackle the problems of heterogeneity and dynamic behavior. Heterogeneity refers to how requests with different QoSs are handled, while dynamic behavior refers to how changes in QoS and changes in multicast group membership are handled. The best scheme to manage VCs should use a minimal number of VCs, while wasting minimal bandwidth due to duplicate packets, and handling heterogeneity and dynamic behavior in a flexible manner [4]. Proposals that significantly alter RSVP should be avoided. Also using special servers might introduce additional delays, so cut-through forwarding approaches are preferred.

The problem of mapping RSVP to ATM is simplified by the fact that while RSVP reservation (RESV) requests are generated at the receiver, actual allocation of resources occurs at the sub-net sender. Thus senders establish all QoS VCs and receivers must be able to accept incoming QoS VCs. The key issues that [4] attempts to tackle are data distribution, receiver transitions, end-point identification and heterogeneity. Several heterogeneity models are defined that provide different capabilities to handle the heterogeneity problem. The dynamic QoS problem can be solved by establishing a new VC, but a timer can be implemented to guarantee that the rate at which VCs are established is not excessively high [4].

## 5.4 Resource Allocation and Admission Control

This section overviews a number of resource allocation and connection admission control mechanisms for multicast connections, other than the allocation mechanisms previously mentioned. RSVP only provides mechanisms for resource allocation in multicast trees, but policies for resource allocation need to be provided.

Connection admission control and resource reservation are complicated for multicast connections. When different receivers have different QoS requirements, most schemes reserve to satisfy the most stringent requirements. Some schemes attempt to later reclaim unneeded resources, while others make use of hierarchical encoding and similar techniques to provide different receivers with different QoS.

In [17], admission control is accomplished through the following steps: first, the end-to-end QoS requirements are divided into local QoS requirements; then, the local QoS requirements are mapped into resource requirements; and, finally, the resources allocated in excess are reclaimed.

An allocation phase initially determines whether there are sufficient resources along the paths to guarantee the QoS delay and loss requirements. A preliminary allocation is then performed. Later, some of the allocated resources are released by taking advantage of sit-uations where different destinations share a path segment and require different amounts of resources on that segment.

Generalized processor sharing is the scheduling mechanism used. Two division policies can be used: even division, and proportional division where fewer resources are allocated on bottlenecked links. Enough resources are reserved to accommodate the tightest QoS achievable at each link of the multicast tree. A link shared by multiple sender-receiver paths is assigned the tightest local QoS requirement. The resources allocated in excess on the multicast tree are later reclaimed without interfering with user traffic [17].

Coders can be used to react to different bandwidth allocations on different branches. A mechanism for feedback control for multicast video distribution over IP was proposed in [6]. The mechanism separates the congestion signal from the congestion control algorithm, so as to cope with heterogeneous networks. The mechanism solicits feedback information in a scalable manner, estimating the number of receivers. The video coder uses the feedback information to adjust its output rate.

Resource allocation mechanisms can be sender-based, rather than receiver based. In this case, reservation mechanisms can exploit known relationships between related connections to allow network resources to be shared between them without sacrificing well-defined guarantees. The network client specifies how traffic from related connections is multiplexed. Unlike RSVP, it is the sender and not the receiver that determines the reservation level. Such protocols are especially useful in cases like conference calls where the relationship between connections is measured and utilized. Such protocols can also protect against unrelated traffic [27].

# 6 Reliable Transport Protocols

Developing a reliable transport protocol for multicast connections has been an active research area in the past few years. The toughest problems in devising a reliable transport protocol for multicast connections include:

- The implosion problem for acknowledgments (ACKs) (or negative acknowledgments NAKs, if used)
- Computing the correct timeout values
- Ensuring fairness
- Multicast group and address maintenance
- Routing support

This section surveys a number of proposals for reliable transport protocols and compares the techniques each of them uses to address the above mentioned problems [9, 30].

The Reliable Multicast Transport Protocol (RMTP) provides sequenced lossless delivery using a selective repeat mechanism. It solves the ACK implosion problem as follows. ACKs are handled based upon a multi-level hierarchical approach. There is a hierarchical tree of designated receivers that cache data, handle retransmission and interact with the ultimate receivers. Designated receivers propagate ACKs up the tree, thus avoiding the ACK implosion problem. Congestion avoidance is implemented using Van Jacobson's slow start algorithm. RMTP allows receivers to join at any point during the connection lifetime [35].

Another protocol to ensure reliable transmission for multicast connections is Reliable Adaptive Multicast Protocol (RAMP) [32, 9]. The protocol is transport-layer reliable (both sender-reliable and receiver-reliable) and adaptive. It uses immediate (not delayed) receiver-initiated, NAK-based, unicast error notification combined with originator based unicast retransmission. This eliminates unnecessary receiver processing overhead, and reduces latency and likelihood of buffer overflow. RAMP is proposed to be useful in ATM networks because the source of packet loss in ATM networks is more likely to be caused by receiver errors and buffer overflows [32].

Computing the appropriate timeout value in a scalable manner can be a challenging problem for multicast connections, where the round trip times to various leaves are different. The optimal timeouts should be computed for each receiver in a multicast tree as a function of the tree topology and the sender-to-receiver delays. The deterministic timeouts for reliable multicast (DTRM) scheme attempts to handle the timeout computation problem, also tackling the NAK implosion problem. The protocol is distributed, sends a single NAK per loss if delay jitter is bounded, and attempts to maximize efficiency by computing timeouts that are optimal with regard to that transport layer window size. It is also end-to-end (switches do not need to send or merge NAKs), and hence, it can be ATM-compatible [26].

Another transport protocol that focuses on the max-min allocation of bandwidth, while tackling the implosion problem is [37]. The protocol allows concurrent and reliable many-to-many multicast, which uses a window-based virtual ring flow control mechanism. A single and immediate acknowledgment message is returned to the sender. Each sender in the group has a single timer, and nodes can join and leave the group dynamically. Bandwidth allocation is max-min fair: the protocol maximizes the bandwidth allocation of each virtual ring, subject to the constraint that an incremental increase in the bandwidth of the ring does not cause a decrease in the bandwidth of another virtual ring whose bandwidth is no more than the initial ring. The algorithm operates by identifying the bottleneck link and dividing the unassigned capacity by the number of virtual rings with unassigned capacity sharing the link, subtracting that capacity, and so on [37].

# 7 Interoperability

The mechanism of migration to a scheme is a very important aspect in the proposal of any new scheme. Hosts or routers implementing the new scheme must be able to transparently interoperate with the components that do not implement the new scheme, so that the new scheme can be gradually deployed. For instance, with IP over ATM, tunneling is used to allow interoperation of components. The interoperability issues are also clearly addressed in RSVP, the SMART scheme and the SEAM scheme. For example, a SEAM based environment can co-exist with islands of non-SEAMable switches, such that only the boundary SEAM switches need to be concerned with interoperating with non-SEAMable islands. Such islands can exploit the point-to-multipoint capabilities of current ATM signaling.

# 8  Open Issues

More extensive studies must be conducted to provide multicast services, either directly over ATM, or on IP running over ATM. We believe that in order to define a truly powerful and flexible ATM multipoint capability, the following problems must be tackled:

1. Comprehensively analyzing the performance of ATM multipoint traffic management under a large variety of configurations and traffic patterns, and using an extremely large number of end systems. Worst case analysis needs to be performed, ensuring scalability requirements are met. Buffer requirements need to be studied, in addition to transient performance, noise, and delays. Both ABR and UBR performance need to be studied.

2. Developing a point-to-multipoint traffic management framework that resolves the consolidation noise and slow transient response issues and balances this tradeoff. The scheme must also have a low overhead and complexity, and give efficient and fair bandwidth allocations.

3. Developing a precise definition of the optimal allocations in a multipoint-to-point connection, and developing a traffic management framework for managing bandwidth for those connections. The scheme must achieve a set of objectives, including optimality of allocations and low overhead.

4. Developing a traffic management framework for multipoint-to-multipoint connections that performs well for all varieties of configurations and traffic patterns, and scales well to a large number of senders and a large number of receivers.

5. Examining the effect of ABR source parameters and ABR source rules on multipoint connections and developing formulae and guidelines for setting for these parameters to achieve the best performance, while maintaining scalability.

6. Allowing heterogeneity and dynamic behavior for multipoint ATM connections. The receivers in a multipoint connection will be allowed to specify different quality of service requirements, and change these requirements dynamically. In addition, best effort traffic can also make use of heterogeneous connections to achieve high utilizations.

7. Developing a signaling, cell forwarding and routing framework for a true ATM multipoint-to-multipoint service, by developing an architecture that handles the cell interleaving, routing, distribution and signaling problems with minimum overhead. The most important consideration for such a scheme is its scalability.

# 9  Summary

The main issues discussed in this paper include:

1. *Signaling and routing issues for multipoint-to-multipoint connections*, and the cell interleaving problem (serialization of packets) need further analysis. The ultimate goal is to define a true ATM multipoint-to-multipoint service that is both simple and scalable.

2. *ATM traffic management for multicast connections*

    - The definition of max-min fairness for multipoint connections is an extension of the definition for point-to-point connections.
    - Switch algorithm extension frameworks proposed for multipoint congestion control must preserve the efficiency and fairness properties of the original point-to-point switch scheme employed in the framework.
    - The main problem specific to point-to-multipoint connections is the consolidation noise problem. This problem occurs when there are distant bottlenecks, and feedback from those bottlenecks is not always received in a timely fashion. Alleviating the consolidation noise problem may create additional problems, such as slow transient response and additional complexity.
    - Transient queues can be avoided by setting the RIF ABR source parameter to a small value, and initial rate overallocation can be overcome by setting the ICR parameter to small values.
    - Multipoint-to-point and multipoint-to-multipoint need further examination.
    - UBR performance must be compared to ABR performance.

3. *Resource allocation and QoS requirements*

    - RSVP performs resource reservations, and should interact with the scheduler to ensure QoS requirements are met.
    - RSVP allows receivers to specify different QoS requirements.
    - RSVP over ATM involves mapping of QoS parameters and setting up the appropriate VCs. The main problems are heterogeneity and dynamic renegotiation of QoS.
    - Usually the most stringent QoS requirements are allocated, excess resources can be reclaimed.

- Ideas from the hierarchical encoding and translators can be borrowed.
- Resource sharing among senders can be used for some applications.

4. *IP over ATM*
   IGMP and mroute are used in IP multicasting. MARS and MCS are proposed for IP over ATM. Several routing issues come into play.

5. *Reliable transport protocols*
   The main problems addressed are:
   - ACK and NACK implosion.
   - Computing the timeout values.
   - Ensuring fairness.

6. *Interoperability issues* are extremely important.